\documentclass[preprint]{aastex}
\def\be{\begin{equation}}
\def\ee{\end{equation}}

\def\HII{\ion{H}{2}}
\def\HA{H$\alpha$}
\def\um{$\mu$m}
\begin{document}
\title{Supernova Remnants and Star Formation in the Large 
Magellanic Cloud}
%
%======================================================================
% authors
%======================================================================
%
\author{Karna M.\ Desai\altaffilmark{1}, You-Hua Chu\altaffilmark{1},
Robert A.\ Gruendl\altaffilmark{1}, William Dluger\altaffilmark{1},
Marshall Katz\altaffilmark{1},
Tony Wong\altaffilmark{1}, C.-H.\ Rosie Chen\altaffilmark{2},
Leslie W.\ Looney\altaffilmark{1}, Annie Hughes\altaffilmark{3,4},
Erik Muller\altaffilmark{5}, J\"urgen Ott\altaffilmark{6}, 
Jorge L.\ Pineda\altaffilmark{7,8}}
\altaffiltext{1}{\itshape Department of Astronomy, University of Illinois
at Urbana-Champaign, 1002 West Green Street, Urbana, IL 61801, USA}
\altaffiltext{2}{\itshape Department of Astronomy, University of Virginia, 
  Charlottesville, VA 22904, USA}
\altaffiltext{3}{\itshape Centre for Supercomputing and Astrophysics, 
  Swinburne University of Technology, Hawthorn VIC 3122, Australia}
\altaffiltext{4}{\itshape CSIRO Australia Telescope National Facility, 
  P.O.\ Box 76, Epping NSW 1710, Australia}
\altaffiltext{5}{\itshape Department of Physics and Astrophysics, 
  Nagoya University, Chikusa-ku, Nagoya 464-8602, Japan}
\altaffiltext{6}{\itshape National Radio Astronomy Observatory, P.O.\ 
 Box O, Socorro, NM 87801, USA}
\altaffiltext{7}{\itshape NASA Postdoctoral Program Fellow}
\altaffiltext{8}{\itshape Jet Propulsion Laboratory, California Institute 
 of Technology, 4800 Oak Grove Drive, Pasadena, CA 91109-8099, USA}
%
%======================================================================
%\begin{document}
%======================================================================
%
%======================================================================
% abstract
%======================================================================
%
\begin{abstract}
It has often been suggested that supernova remnants (SNRs) can
trigger star formation.  
To investigate the relationship between SNRs and star formation,
we have examined the known sample of 45 SNRs in the Large Magellanic 
Cloud to search for associated young stellar objects (YSOs)
and molecular clouds.  We find seven SNRs associated with both 
YSOs and molecular clouds, three SNRs associated with YSOs but 
not molecular clouds, and eight SNRs near molecular clouds 
but not associated with YSOs. 
Among the 10 SNRs associated with YSOs, the association between 
the YSOs and SNRs can be either rejected or cannot be convincingly
established for eight cases.  Only two SNRs have YSOs closely aligned 
along their rims; however, the time elapsed since the SNR began to
interact with the YSOs' natal clouds is much shorter than the
contraction timescales of the YSOs, and thus we do not see any 
evidence of SNR-triggered star formation in the LMC.
The 15 SNRs that are near molecular clouds
may trigger star formation in the future when the SNR shocks have 
slowed down to $<$45 km~s$^{-1}$.  We discuss how SNRs can alter 
the physical properties and abundances of YSOs.
\end{abstract}

\subjectheadings{Supernova Remnants; stars: formation; Magellanic Clouds}

\maketitle
\newpage
%
%======================================================================
\section{Introduction}  \label{sec:intro}
%======================================================================
%
Star formation is frequently seen near supernova remnants (SNRs).
For example, young stars have been detected in the vicinity 
of the Galactic SNR G54.1+0.3 \citep{Ketal08}, an SNR near the giant
\HII\ region N66 in the Small Magellanic Cloud \citep{Getal08}, 
and two SNRs in the Large Magellanic Cloud \citep[LMC;][]{CG08}.
Even the solar system appears to have formed at a site where a recent
supernova explosion had occurred, as old meteorites show an anomalously 
high $^{60}$Ni abundance that requires short-lived radioactive $^{60}$Fe 
produced by core-collapse supernovae be injected into the nascent solar 
nebula \citep{VB02,Metal05,Tetal06}.

The observed physical association between star formation and SNRs does 
not necessarily imply a causal relationship for two reasons.
First, most massive stars are formed in clusters or OB associations 
where star formation may continue and propagate outward for a prolonged 
period of time; thus, SNRs produced by core-collapse supernovae are
likely to be near young stars in star-forming environments.
Second, theoretical calculations show that shocks at velocities of
20--45 km~s$^{-1}$ may compress molecular clouds to trigger star
formation, while faster shocks will destroy molecular clouds
\citep{VC98}.  SNRs are confirmed by their simultaneous exhibition 
of diffuse X-ray emission, nonthermal radio spectral index, and 
high [\ion{S}{2}]/\HA\ ratios \citep[e.g.,][]{Metal83}, and these
signatures are produced by high-velocity shocks.  For example, 
X-ray-emitting plasma at temperatures $\ge$10$^6$ K can be produced 
in adiabatic shocks only if the shock velocity is $\ge$300 km s$^{-1}$.
Therefore, confirmed SNRs have strong shocks that are destructive 
to molecular clouds.
%while evolved slowly-expanding SNRs that can compress ambient 
%molecular clouds to trigger star formation are no longer 
%identifiable as SNRs.  
It has thus been suggested that the star formation observed near 
SNRs was triggered by the expansion of \HII\ regions or wind-blown 
bubbles of the supernovae's massive progenitor
stars \citep{Ketal08,Getal08}.  This hypothesis is reasonable because 
small wind-blown bubbles with expansion velocities of 15--25 km s$^{-1}$
are commonly observed in \HII\ regions associated with young OB 
associations \citep{Netal01} and large superbubbles formed by OB 
associations have typical expansion velocities of 20--45 km s$^{-1}$
\citep{Detal01}.

We have used the sample of known SNRs in the LMC to make a broad 
investigation of the relationship between SNRs and star formation.
The LMC sample of SNRs was initially compiled by \citet{Metal83,Metal84,
Metal85} using \emph{Einstein} X-ray sources and radio observations.
It has grown with additional members diagnosed by \emph{ROSAT} 
X-ray observations \citep{Cetal93,Cetal95,Cetal97,Setal94} or by
high [\ion{S}{2}]/\HA\ ratios \citep{Petal04}.
The 45 currently known LMC SNRs are listed in Table 1.
Recent star formation is best depicted by the presence of young 
stellar objects (YSOs).  The LMC YSOs have been identified
and inventoried by \citet[][hereafter GC09]{GC09} and \citet{Wetal08} using 
\emph{Spitzer Space Telescope} observations.  
As discussed in detail by GC09, the YSO catalog of \citet{Wetal08} 
suffers problems of incompleteness and contamination by background 
galaxies, and therefore will not be used in this work.
We have combined the LMC SNRs in Table 1 and the GC09 YSO catalog to 
search for star formation in juxtaposition with SNRs.  We find 10 SNRs 
associated with YSOs, and 7 SNRs appear to be associated with molecular 
clouds but do not have YSOs.
In this paper we describe the data sets and the search method in section 2, 
discuss star formation in the vicinity of SNRs in section 3, and discuss 
the implications of this study in section 4.

%
%======================================================================
\section{Data Sets and Methodology}
%======================================================================
%
To identify the location of SNRs in the LMC, we use \HA, [\ion{O}{3}],
and [\ion{S}{2}] images from the Magellanic Cloud Emission-Line
Survey \citep[MCELS;][]{Setal99}.  The MCELS images were taken with
the Curtis Schmidt Telescope at the Cerro Tololo Inter-American
Observatory (CTIO).  The individual images have a pixel size of
2\farcs3~pixel$^{-1}$, but the mosaicked images have a pixel size of 3\arcsec.
The [\ion{O}{3}] filter includes only the $\lambda$5007 line, while
the \HA\ filter includes the \HA\ $\lambda$6563 and [\ion{N}{2}] 
$\lambda\lambda$6548, 6584 lines, and the [\ion{S}{2}] filter includes
both the $\lambda$6716 and $\lambda$6731 lines.  Thus, the MCELS \HA\ 
images are really \HA+[\ion{N}{2}] images, although the [\ion{N}{2}] lines
are usually less than 10\% as strong as the \HA\ line because of
the low nitrogen abundance of the LMC \citep{RD90}.
In general, the H$\alpha$ images are best in showing both collisionally 
ionized gas in SNRs and photoionized gas in \HII\ regions; the 
[\ion{O}{3}] images show the cooling
zone closest to SNR shock fronts and the gas photoionized by early-type O
stars; and the [\ion{S}{2}] images best delineate the boundaries of 
SNRs interacting with a dense medium.  Young Type Ia SNRs in a
mostly neutral medium are dominated by hydrogen Balmer lines, and hence
visible only in the H$\alpha$ images \citep{CKR80}.

We adopt the GC09 list of YSOs in the LMC.
These YSOs, limited by the sensitivity and resolution of
{\it Spitzer} observations, have masses greater than $\sim$4 $M_\odot$
and correspond to intermediate- and high-mass stars.
To examine the location of YSOs relative to the SNRs, we have
used archival \emph{Spitzer Space Telescope} InfraRed Array 
Camera \citep[IRAC;][]{Fazio04} images in the
3.6, 4.5, 5.8, and 8.0 $\mu$m bands and Multiband Imaging
Photometer for {\it Spitzer} \citep[MIPS;][]{Rieke04} images in 
the 24 and 70 $\mu$m bands, as used in GC09.
The IRAC images have a resolution of
$\sim$2$''$, and the MIPS images have a resolution of $\sim$6$''$
at 24 $\mu$m and $\sim$17$''$ at 70 $\mu$m. The IRAC 8.0 $\mu$m 
images are useful in showing the polycyclic aromatic hydrocarbon 
(PAH) emission from partially ionized regions. They are also 
best for illustrating the locations of YSOs because of the
combination of long wavelength and high angular resolution.

To examine molecular clouds associated with SNRs and star
formation, we have used the CO (J=1$\rightarrow$0) observations from the
NANTEN survey \citep{Fetal01,Fetal08} and Magellanic Mopra Assessment 
(MAGMA) survey \citep{Oetal08,Hetal10}.  The NANTEN observations
were made with a 4-m radio telescope with a 2\farcm6 half-power beam
and a velocity resolution of 0.65 km s$^{-1}$; this survey 
covered the entire LMC with a grid spacing of 2$'$.
The 3$\sigma$ detection limit of the NANTEN survey corresponds to an
integrated CO intensity of $I_{\rm CO}$ = 1.2 K km s$^{-1}$ or $N({\rm H}_2)
\sim 8 \times 10^{21}$ cm$^{-2}$ using a conversion factor of $X_{\rm CO} =
7 \times 10^{20}$ cm$^{-2}$ (K km s$^{-1}$)$^{-1}$.  The NANTEN
survey is estimated to be complete for molecular clouds with 
masses down to $5 \times 10^4$ $M_\odot$.
All known SNRs in the LMC have been covered by the NANTEN survey.
The MAGMA observations were made with the Mopra 22-m telescope 
with an angular resolution of 33$''$ and a velocity resolution of 
0.16 km s$^{-1}$. The MAGMA survey covered only the $\sim$100 
brightest giant molecular clouds (GMCs) that had been previously 
detected by the NANTEN survey; thus, not every LMC SNR has MAGMA
observations available.  
In this paper we have used integrated
intensity maps from MAGMA smoothed to 45$''$ resolution.
An estimate of the MAGMA detection limit is $I_{\rm CO}$ = 2 K km
s$^{-1}$ or $N({\rm H}_2) \sim 1.4 \times 10^{22}$ cm$^{-2}$ using the 
same conversion factor as used by \citet{Fetal08}.  Since this is 
averaged over a 45$''$ beam, most (if not all) clouds with emitting 
structures smaller than the NANTEN beam will be detected by MAGMA, 
but smoothly distributed emission that is just barely detected by 
NANTEN may escape detection in the MAGMA data.  The range of cloud
masses detected by MAGMA is 1.6$\times$10$^4$ -- 6$\times$10$^6$ 
$M_\odot$ \citep{Hetal10}.  Small molecular clouds or low-density 
molecular gas associated with diffuse 8 $\mu$m emission would not 
be detected by either survey.

The 45 LMC SNRs that we have examined are listed in Table 1.
The coordinates and sizes of SNRs are measured from the MCELS 
images.  Some SNRs are better defined in {\it Chandra} X-ray or 
{\it Spitzer} 24 $\mu$m images, then these images are used
to measure the SNRs' positions and sizes.
We made a set of figures to identify the optical boundaries
of the SNRs and compare them with the locations of YSOs
and molecular clouds (a subset of these figures are presented
in Figures 1 and 2).  For each SNR, the figure contains three
panels showing MCELS H$\alpha$, MCELS [\ion{S}{2}], and {\it Spitzer}
IRAC 8.0 $\mu$m images with an identical field-of-view. 
%15$'$$\times$15$'$ field-of-view.  
Although SNRs are diagnosed by their higher [\ion{S}{2}]/H$\alpha$ 
ratios, the ratios are still $<$1.0 and most SNRs are still 
detected with the highest signal-to-noise ratios in H$\alpha$ 
images.  YSOs are enshrouded by dust and thus appear as point 
sources in the 8.0 $\mu$m images.  The YSOs identified 
by GC09 are marked on the H$\alpha$ and 8.0 $\mu$m images.  
To show the distribution of molecular clouds, we have plotted the 
NANTEN and MAGMA CO contours over the H$\alpha$ and 8.0 $\mu$m 
images, respectively.  The contour levels follow a geometric 
progression starting at the 3$\sigma$ level (i.e., 3$\sigma$, 
6$\sigma$, 12$\sigma$, 24$\sigma$, ...), where $\sigma$ is the
typical root-mean-square noise level of the integrated intensity maps.

The positions of the YSOs have been marked with different
colors and symbols depending on GC09's assessment of their nature.  
The YSOs marked in red have [8.0]$<$8.0 mag while those marked in 
green have [8.0]$>$8.0 mag, roughly corresponding to massive 
($\gtrsim$ 10 $M_\odot$) and intermediate-mass ($\sim$ 4--10 $M_\odot$)
stellar cores, respectively.
The symbols used to mark the locations of the YSOs reflect the relative
certainty that they are bona-fide YSOs. Open circles mark YSOs
classified as ``Definite YSOs''  or ``Probable YSOs,'' and
crosses mark ``Possible YSOs.''  The ``Possible YSOs''
are more likely to be non-YSOs than YSOs, but they can not be
categorically ruled out that they are YSOs.
 
We have examined the figures carefully, comparing the relative
locations of SNRs, YSOs, and molecular clouds.
Since the LMC is a nearly face-on disk galaxy 
\citep[i=35$^\circ$;][]{vdMC01}, it is reasonable to
assume association between two objects
whose projected boundaries overlap.
In our initial inspection, SNRs with YSOs projected within
or in close vicinity are selected for further investigation.
We use the NANTEN 3$\sigma$ contours for the boundary 
of molecular clouds and if the optical boundary of an SNR overlaps
a molecular cloud we assume they are associated.
This initial examination finds 3 SNRs associated with YSOs but
not molecular clouds, 7 SNRs associated with both YSOs and 
molecular clouds, and 8 SNRs associated with molecular clouds
but not YSOs, as listed in the last two columns in Table 1.
The former 10 SNRs associated with YSOs are presented in 
Figure 1, and the latter 8 SNRs associated with molecular 
clouds in Figure 2. The associations between these SNRs 
and YSOs or molecular clouds are further evaluated using
higher-resolution maps, such as MAGMA and ESO-SEST, and 
discussed in detail in the next section.
%
%======================================================================
\section{Star Formation in the Vicinity of SNRs}   % Results
%======================================================================
%

In this section we will first discuss the SNRs associated with YSOs,
describing the relative locations of SNRs, YSOs, and
molecular clouds. We further examine the relationships among YSOs, 
SNRs, and  ionization fronts to probe possible triggering mechanisms for
the star formation.  Then we discuss the SNRs projected near 
molecular clouds but showing no star formation, and examine whether the 
SNRs are physically interacting with the molecular clouds.

%
%------------------------------------------------------------------------
\subsection{SNRs Associated with YSOs}
%------------------------------------------------------------------------
%

\noindent {\bf SNR J0449.3$-$6920 (MCELS J0449-6921)}

This SNR has a diameter of 2$'$ (30 pc). The SNR shell,
best seen in the [\ion{S}{2}] image, is well defined on
the north and south sides.  The \HA\ image shows
that the SNR blends with \HII\ regions to its west 
and east. Two YSOs are present in the vicinity of the SNR.
The \HA\ image shows that one YSO is located in a dark 
cloud to the east of the eastern \HII\ region, and hence 
it is not associated with the SNR.  
The other YSO is located on the eastern rim of
the western \HII\ region. As the YSO is located in
the partially ionized surface of the western diffuse \HII\
region (the large elliptical shell structure in the 
8.0 \um\ image), it is most likely that its formation
was triggered by the expansion of this western diffuse \HII\   
region. No molecular clouds in the vicinity of the SNR were
detected by NANTEN.

\noindent {\bf SNR J0455.6$-$6838 (N86)}

This SNR has a multiple-lobe structure, with the longest dimension
approaching 100 pc \citep{Wetal99}.
The northwest corner of the SNR borders the southern extension of
the NANTEN GMC J0455$-$6830 \citep{Fetal08}. The MAGMA observations
covered the core of this GMC but not the southern extension.
A YSO is detected near the border between the SNR and the GMC. 
As the exact boundary of the SNR is not known, it is not clear 
whether the YSO is inside the SNR boundary and physically
associated, or outside the boundary and unrelated to the SNR.

\noindent {\bf SNR J0513.2$-$6912 (SNR 0513$-$692)}

This SNR is 4\farcm5$\times$3\farcm2 (68 pc $\times$ 48 pc)
in size.  The northeastern quadrant of the SNR overlaps
a small molecular cloud that was detected but not 
cataloged by NANTEN \citep{Fetal08}.  No MAGMA observations
are available for this small cloud.  In the region where the 
SNR overlaps the molecular cloud, there exist a small dense 
\HII\ region and three YSOs. The 8.0 \um\ image shows that
the YSOs are located along filaments that extend to the 
northeast and belong to a larger structure unrelated to 
the SNR.  

To show the detailed relationship among the 
small \HII\ region, SNR, YSOs, and the 8.0 \um\ filaments,
we present in Figure 3 a close-up, high-resolution \HA\ image 
obtained with the MOSAIC camera on the CTIO 4m telescope 
(from GC09), along with the 8.0 \um\ image.  The \HA\ image 
shows that the massive YSO is coincident with the brightest 
\HA\ knot, which 
is separated from the rest of the \HII\ region by a dust lane.  
One intermediate-mass YSO is located in the southwest extension 
of the dust lane, while the other intermediate-mass YSO is 
located at 5 pc from the southern edge of the \HII\ region.

The close alignment of the three YSOs along the SNR rim may
lead to the impression that the SNR triggered the formation of
these YSOs; however, this speculation is not supported by the
dynamic timescale of the SNR and the ages of the YSOs. 
The SNR J0513.2$-$6912 has an average radius of 30 pc.
Although its expansion velocity ($V_{\rm exp}$) is unknown, we 
may adopt the typical values of 150--300 km s$^{-1}$ observed 
in confirmed SNRs in the Magellanic Clouds \citep{CK88}.
Assuming a Sedov phase for the SNR, its dynamic age would be
4--8$\times$10$^4$ yr.  The interaction of the SNR and the
molecular cloud occurred only during the last $\sim$20\% of
its lifetime, i.e., 0.8--1.6$\times10^4$ yr.  This timescale
is much shorter than the contraction timescale of 
intermediate-mass stars, $\sim10^6$ yr, and massive stars,
$\sim10^5$ yr \citep{BM96}.  Thus, it is impossible for the 
SNR to have triggered the formation of these YSOs.

One possible cause for the close alignment of YSOs and SNR rim
is that the supernova exploded inside an interstellar bubble
blown by its progenitor massive star.  Interstellar bubbles 
have mild expansion velocities, 15--25 km s$^{-1}$ \citep{Netal01},
and may trigger star formation.  In this case, the supernova
ejecta initially expands rapidly in the low-density interior 
of the bubble and slows down when it encounters the dense 
bubble shell; thus, the SNR boundary is dictated by the
previous interstellar bubble and the YSOs triggered by the
interstellar bubble naturally appear along the SNR shell rim.

\noindent {\bf SNR J0532.5$-$6731 (SNR 0532$-$675 in N57)}

This SNR is in the northern portion of the superbubble
in LHA120-N57 \citep{Henize56}.
It was first suggested by \citet{Metal85}, and is best
shown in recent {\it XMM-Newton} observations 
(Maddox et al.\ 2010, in preparation).
X-ray contours extracted from these X-ray observations are
overplotted on the [\ion{S}{2}] image to show the boundary 
of the SNR, $\sim$4\farcm5 in diameter.  The NANTEN GMC 
J0532$-$6730 is detected on the northeast side of the SNR,
but the MAGMA map shows no overlap between the GMC and the SNR.
One YSO is detected within the GMC,
but it is projected outside the boundary of the SNR, indicating 
that the SNR and its progenitor are not responsible for the 
formation of this YSO.

\noindent {\bf SNR J0535.7$-$6602 (N63A)}

The SNR is expanding within a bubble produced by its progenitor 
within the N63 \HII\ complex.  X-ray contours extracted
from {\it Chandra} observations \citep{Wetal03} are overplotted 
on the [\ion{S}{2}] image to show the boundary of the SNR, 
$\sim$1\farcm2 in diameter.  No
molecular clouds are detected by NANTEN.  Two YSOs are 
projected within the boundary of N63, but well outside the SNR.
It is likely that the expansion of the N63 \HII\ region triggered 
the formation of these YSOs. This suggestion is further supported 
by the location of one YSO in the partially ionized eastern rim 
of the N63 \HII\ region (see the 8.0 \um\ image in Figure 1). 
A detailed study of star formation in the \HII\ region N63
has been reported by \citet{Cetal08}.

\noindent {\bf SNR J0536.0$-$6735 (DEM\,L241)}

This SNR has a shell like structure and is associated with the
OB association LH88 \citep{LH70}; it has a relatively small 
expansion velocity, $\sim$70 km~s$^{-1}$, suggesting that the SNR 
may be expanding into a stellar-wind blown bubble \citep{Chu97}.
The SNR overlaps the NANTEN GMC J0535$-$6735.
This GMC is resolved by MAGMA observations into clumps,
and the two largest clumps contain multiple YSOs. 
These molecular clumps and the YSOs are unrelated to the SNR.
One YSO is projected within the SNR, but as this SNR is superposed
on an OB association, it is uncertain whether this YSO is truly 
associated with this SNR and its progenitor or with other members
of LH88.  

\noindent {\bf SNR J0537.5$-$6627 (DEM\,L256)}

This SNR has also been identified as MCELS J0537$-$6628.
The northern part of this SNR overlaps the NANTEN GMC 
J0537$-$6626.  One massive and two intermediate-mass
YSOs are detected on the northwest rim of the SNR
within the region overlapping the GMC. The 8.0 \um\ image
shows that the three YSOs are aligned along an arc of PAH 
emission.  To provide a detailed view of this region, we
present a high-resolution H$\alpha$ close-up image, 
along with the 8 \um\ image, in Figure 4.  We have also
analyzed the underlying stellar population using the 
Magellanic Cloud Photometric Survey \citep{Zetal04}, and
identified one star within the mass range 20--25 $M_\odot$
($\sim$O8 spectral type, marked with a blue triangle in 
Figure 4) and two stars within 15--20 $M_\odot$ ($\sim$B0
spectral type, marked with orange triangles in Figure 4).
The two less massive stars are inside small \HII\ regions
and one is coincident with an intermediate-mass YSO.
While these YSOs and early-B stars may contribute to the 
excitation of PAHs along the 8 \um\ arc, the surface
brightness of the 8 \um\ arc suggests nearly
uniform illumination/excitation, which is more
naturally provided by the O8 star at a distance (near
the southeast corner of Figure 4).

High-dispersion echelle spectra of this SNR show a non-uniform
expanding shell with the largest expansion velocity occurring
on the receding side and reaching $\sim$120 km s$^{-1}$
\citep{Ketal06}.  Adopting this as the expansion velocity
and assuming a Sedov phase, the dynamic age of the SNR is 
$\sim$8$\times$10$^4$ yr.  If we further assume that the
interaction of SNR and the YSOs' natal cloud occurred during
the last 20\% of the SNR lifetime, there would not be sufficient
time for the clouds to contract and form these YSOs.  Therefore,
similar to SNR J0513.2$-$6912, SNR J0537.5$-$6627 cannot have
triggered the formation of the YSOs along its northwest rim.
Likewise, the supernova most likely exploded in the central 
cavity of a bubble blown by its massive star progenitor, and
the YSOs triggered by the bubble expansion would naturally
fall along the SNR rim.

\noindent {\bf SNR J0537.8$-$6910 (N157B)}

This SNR is in an \HII\ region ionized by the
OB association LH99 \citep{Chu97}.
With a complex nebular background, the SNR boundary is best
determined using nebular kinematics \citep{Cetal92} or diffuse
X-ray emission \citep{Cetal06}. A low-level X-ray contour is 
plotted in Figure 1 to illustrate the rough boundary of this SNR.
The NANTEN survey shows a large molecular complex over the
SNR, but the molecular complex is resolved by MAGMA 
observations into clumps.  No molecular peak is detected 
within the SNR boundary, although numerous dark clouds or 
dusty features are seen.
Two massive and three intermediate-mass YSOs are
detected within the boundary of the SNR, with the two 
massive ones deeply embedded in dark clouds. The complexity 
in this \HII\ region precludes unambiguous associations
 between the YSOs and the SNR. 

\noindent {\bf SNR J0540.0$-$6944 (SNR in N159)}

This SNR is associated with N159 \citep{Cetal97,Wetal00}. 
N159 is an active star forming region
that hosts the OB association LH105. The NANTEN data show 
that N159 is associated with the northern end of the large
molecular ridge of the LMC.  The MAGMA map has resolved 
this region into multiple concentrations \citep{Oetal08} and
shows that the YSOs are projected near the molecular peaks.
This region is too complex to make unambiguous 
association between YSOs and the SNR. A detailed discussion
of the star formation in N159 has been reported by \citet{Cetal10}.

\noindent {\bf SNR J0543.1$-$6858 (DEM\,L299)}

This large SNR shows a central cavity of dimension
1\farcm8 $\times$ 2\farcm3 (27 pc $\times$ 35 pc).
A YSO is projected within a small bubble-like structure, $\sim1'$ 
across, at the northwest rim of the SNR.  The 8.0 \um\ image shows 
that the YSO is located in a ridge of enhanced diffuse emission,
which is most likely PAH emission at the boundary between ionized and
neutral media.  A close-up, high-resolution \HA\ image of this small
\HII\ region is shown in Figure 5, along with an 8.0 \um\ image.
The curved ridge of 8.0 \um\ emission is centered on a bright
star that has been cataloged as Sk$-68$ 155 and classified
as a B0.5 (based on optical data) and O8 Ia (based on UV data)
by \citet{Metal99}.  Using photometric measurements of 
Sk$-$68 155, $U$ = 12.04, $B$ = 12.76, and $V$ = 12.72 \citep{Nico78},
we estimate a spectral type of $\sim$B0 Ia, roughly consistent
with the previous classifications.  With this spectral type, 
Sk$-68$ 155 is able to photoionize this \HII\ region,
excite the PAHs, and possibly trigger the formation of the YSO.
This small \HII\ region and its associated PAH features and YSO
are exterior to the SNR boundary, and thus not triggered by 
SNR J0543.1$-$6858.

%
%--------------------------------------------------------------------------
\subsection{Other SNRs Projected Near Molecular Clouds}  
%-------------------------------------------------------------------------
%

Eight SNRs are projected near molecular clouds, but no YSOs 
are observed within the SNR boundaries. Figure 2 shows images of
these SNRs. Close examinations of the NANTEN and MAGMA contours
in Figure 2 reveal that four SNRs do not seem to be in direct 
contact with molecular clouds: J0518.7$-$6939 in N120, J0521.6$-$6543, 
J0523.1$-$6753 in N44, and J0550.5$-$6823.   
In the case of J0526.0$-$6604 (N49), the SNR overlaps with a small
feature in the MAGMA map, but the feature is smaller than the MAGMA
resolution and the channel maps do not detect this feature in 
any adjacent velocity channels, suggesting that it is just a noise
bump.  A deep ESO-SEST CO J=2$\rightarrow$1 observation shows a small 
cloud, with a mass of 9$\times$10$^3$ $M_\odot$ assuming 
$N_{\rm H_2}$/$W_{\rm CO(2-1)}$ = 1.87$\times$10$^{21}$ cm$^{-2}$ 
(K km s$^{-1}$)$^{-1}$, on the eastern rim of N49 \citep{Betal97}.
This small cloud would be below the detection limit of both
the NANTEN and MAGMA surveys.
In the case of SNR N186D, the large beam size and unavailability
of the channel maps of the NANTEN survey make it impossible to assess 
whether the SNR is truly interacting with the molecular cloud. 

The remaining two SNRs, N103B and N132D, are projected within GMCs
in both the NANTEN and MAGMA surveys.
The SNR N103B has been shown to be the descendant of a Type Ia
supernova based on the abundances derived from X-ray 
observations \citep{Hetal95,vdh02}. 
As the progenitor of the Type Ia supernova is not a massive star
with ionizing power, there was no \HII\ region or wind-blown bubble
to trigger star formation in the GMC; therefore, the lack of 
YSOs around the SNR N103B is expected.
In the case of N132D, high-resolution ESO-SEST observations  
\citep[CO J=2$\rightarrow$1;][]{Betal97} show that the molecular cloud is 
south of the SNR boundary.  The apparent overlap between the 
molecular cloud and N132D in the NANTEN and MAGMA maps is caused
by lower resolution and the smoothing of the data.
Based on the high-resolution ESO-SEST results, we do not see 
clear indication that the SNR or its progenitor has significantly
interacted with the GMC; therefore, it is not surprising that
no YSOs exist around the SNR N132D.

\section{Discussion and Conclusions}

The SNR-triggered star formation mechanism has been of great interest,
but is frequently suggested casually without substantiation.
In order to study the current and possible future star formation
associated with SNRs, we have examined the entire sample of 45 
SNRs in the LMC and assessed their relationship with YSOs and
molecular clouds. 
Initial examinations using only the NANTEN maps of molecular clouds
found seven SNRs associated with both YSOs and molecular clouds,
three SNRs associated with YSOs but not molecular clouds, and
eight SNRs near molecular clouds but not associated with YSOs.
These 18 SNRs were further scrutinized with higher-resolution
molecular maps in conjunction with {\it Spitzer} IRAC 8.0 $\mu$m
images.

Among the 10 SNRs associated with YSOs, two (N157B and the SNR in
N159) are in regions with very active star formation, so it is 
difficult to assess their relationship with the YSOs.
In four cases, either the YSOs are clearly outside the SNR boundary 
(N63A and the SNR in N57), or the association between YSOs and 
the SNRs is uncertain (N86 and DEM\,L241).  In the remaining
four cases, YSOs are projected near the SNR rims.  Upon closer
examination, the YSOs near J0449.3$-$6920 and DEM\,L299 are 
located within 8.0 \um\ filaments associated with neighboring 
\HII\ regions, and hence not triggered by the SNRs.  Only SNRs
J0513.2$-$6912 and DEM\,L256 have YSOs closely aligned along 
their rims; however, the time elapsed since the SNR began to
interact with the YSOs' natal clouds is much shorter than the
contraction timescales of the YSOs, and thus the YSOs could not
have been triggered by the SNRs.  

From the sample of confirmed LMC SNRs, we do not see any evidence 
of SNR-triggered star formation.  
This result is consistent with that of theoretical considerations
of SNR shocks interacting with molecular clouds:
young SNRs with powerful shocks destroy molecular clouds, but older 
SNRs whose shock velocities have slowed down to $<$45 km~s$^{-1}$ may
compress molecular clouds to form stars \citep{VC98}.
However, observationally it is difficult to confirm or refute
SNR-triggered star formation because such old SNRs have already
lost their distinguishing characteristics at radio and X-ray
wavelengths.

%SNR-triggered star formation has been suggested, for example, 
%by \citet{Oetal04} to explain  the low-mass star formation in 
%the $\beta$ Pictoris moving group.
%It is difficult to confirm or refute this mechanism because the
%stars are 10$^6$--10$^7$ yr old and the assumed SNR has long lost
%its distinguishing characteristics, i.e., X-ray emission and
%radio synchrotron radiation.
%However, through modeling of SNR-cloud interactions, it is possible
%to constrain the physical conditions that lead to 
%SNR-triggered star formation.
%For example, a 3D radiatively cooling 
%numerical simulation of SNR-cloud interaction by \citet{Metal06} 
%concludes that SNR-triggered star formation for the $\beta$ 
%Pictoris moving group is implausible under the initial conditions
%specified by \citet{Oetal04}.

The frequent occurrence of star formation near young SNRs 
resulting from core-collapse supernovae should not be
surprising because massive stars tend to form in clustered
environment where expanding \HII\ regions and interstellar
bubbles can trigger further star formation.  In our study of
LMC SNRs, we find at least five SNRs (J0449.3$-$6920, 
J0513.2$-$6912, N63A, DEM\,L256, and DEM\,L299) having YSOs 
located along arcs or filaments of PAH emission in the 8 $\mu$m 
images, suggesting that the star formation was triggered 
by advancing ionization fronts or expanding \HII\ regions.

Fifteen of the SNRs in the LMC are near molecular clouds, and in some
cases the SNRs may have already interacted with the molecular 
clouds, but the fast SNR shocks cannot trigger star formation yet.
As the SNRs age (10$^5$--10$^6$ yr) and their shock velocities
have slowed to $<$45 km~s$^{-1}$, their interactions with the
molecular clouds may trigger star formation.
About 33\% (15/45) of the known SNRs in the LMC have the potential to 
trigger star formation in the future.
It is interesting to note that when SNRs can trigger star formation,
the currently associated massive YSOs would have become more evolved 
main sequence stars and their ionization fronts in the ambient medium
can also trigger further star formation.
Therefore, in a complex environment with continuous star formation, 
it is often impossible to determine exactly the triggering mechanisms 
for star formation.

We have discussed the relationship between SNRs and the current
and future star formation.  
In all cases, SNRs may alter the physical properties of the YSOs.
For YSOs near SNRs, the passage of SNR shocks
may strip the envelopes of YSOs and affect the accretion
of mass.  It would be interesting to examine the spectral
energy distributions of these YSOs and search for differences from
those of normal YSOs.
YSOs near SNRs may intercept supernova ejecta so that 
heavy elements and radioactive material are incorporated into the 
circumstellar envelopes and disks.
This is similar to what \citet{Letal06} have suggested for our solar
nebula.
SNRs interacting with molecular clouds can inject heavy elements into
the clouds for future star formation

It has often been assumed that an aged SNR both triggers
the formation and enriches the circumstellar material of a YSO
\citep[e.g.,][]{Betal10}.
However, our investigation of SNRs and star formation shows
that SNRs with massive star progenitors are likely located near 
molecular clouds and their progenitors' or neighboring massive 
stars' ionization fronts and wind-blown bubbles can trigger star
formation.  It is thus common to have YSOs near an SNR 
to be enriched with the supernova ejecta, but their formation 
was not triggered by the SNR.

Finally, we note that clustered supernovae may produce large
shell structures of sizes $\sim$100 pc, such as superbubbles
blown by OB associations, or $\sim$1000 pc, such as 
supergiant shells produced collectively by multi-generations
of massive star formation \citep{Chu08}.
These large shells have moderate expansion velocities, usually
$\ll$50 km s$^{-1}$, and thus may trigger star formation.
Indeed, star formation has been observed around the rims of 
superbubbles, e.g., N44 \citep{Cetal09}, and supergiant
shells. e.g., LMC-4 \citep{Yetal01,Betal09}.
While supernova explosions have led to the formation of these
shell structures, the star formation is not attributed to
any single SNR and therefore should not be referred to as
``triggered by an SNR''.

\acknowledgments
We thank the referee for his/her thorough reading and useful 
suggestions to improve the paper.   We also thank Dr.\ Y.\ Fukui 
for providing the NANTEN data.
This research was supported by NASA grants JPL 1290956
and JPL 1316421 and NSF grant AST 08-07323.
JLP was supported by an appointment to the NASA Postdoctoral
Program at the Jet Propulsion Laboratory, California Institute 
of Technology, administered by Oak Ridge Associated Universities 
through a contract with NASA.

\eject

\begin{figure}
\plotone{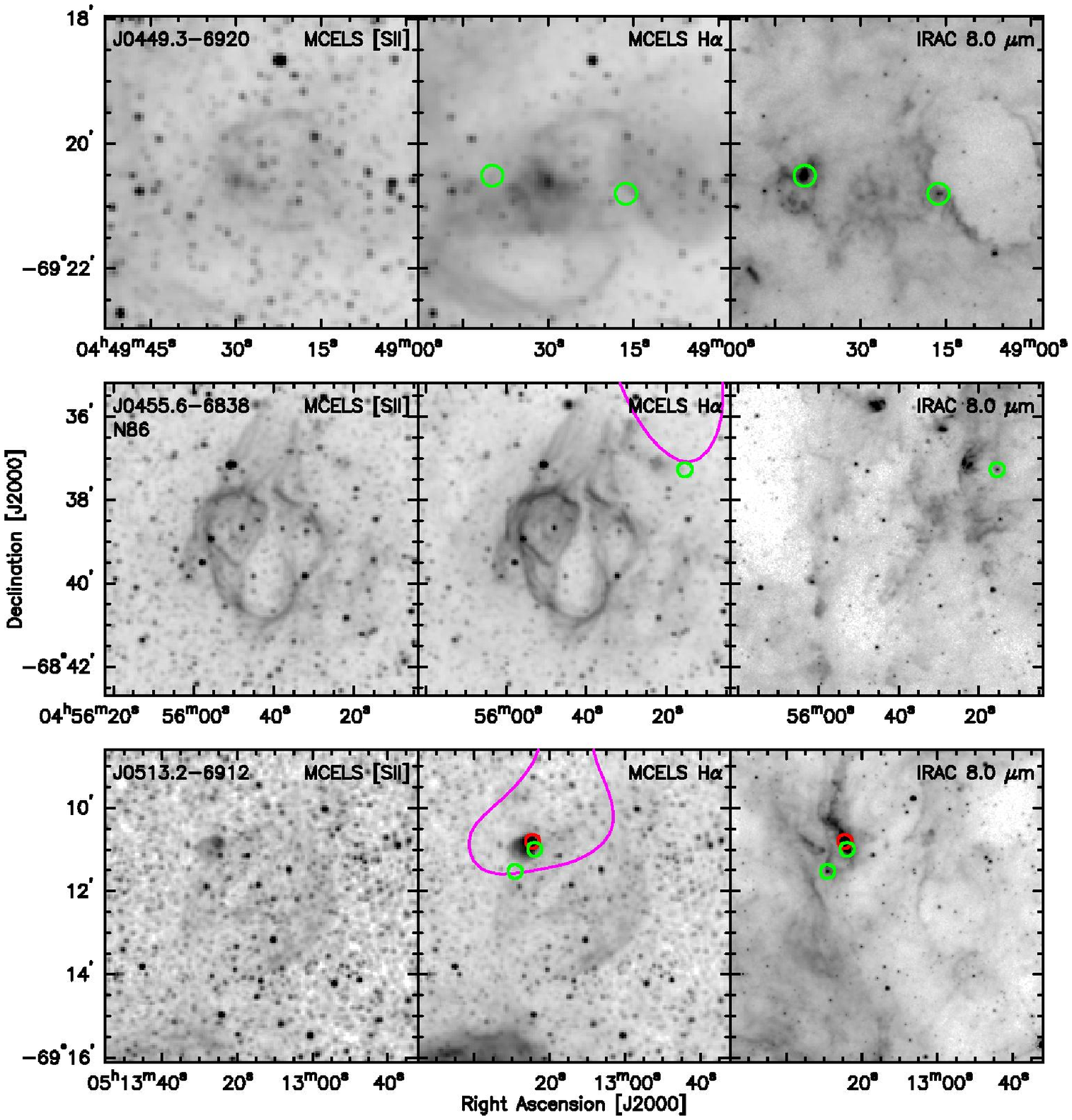}
\end{figure}
\begin{figure}
\plotone{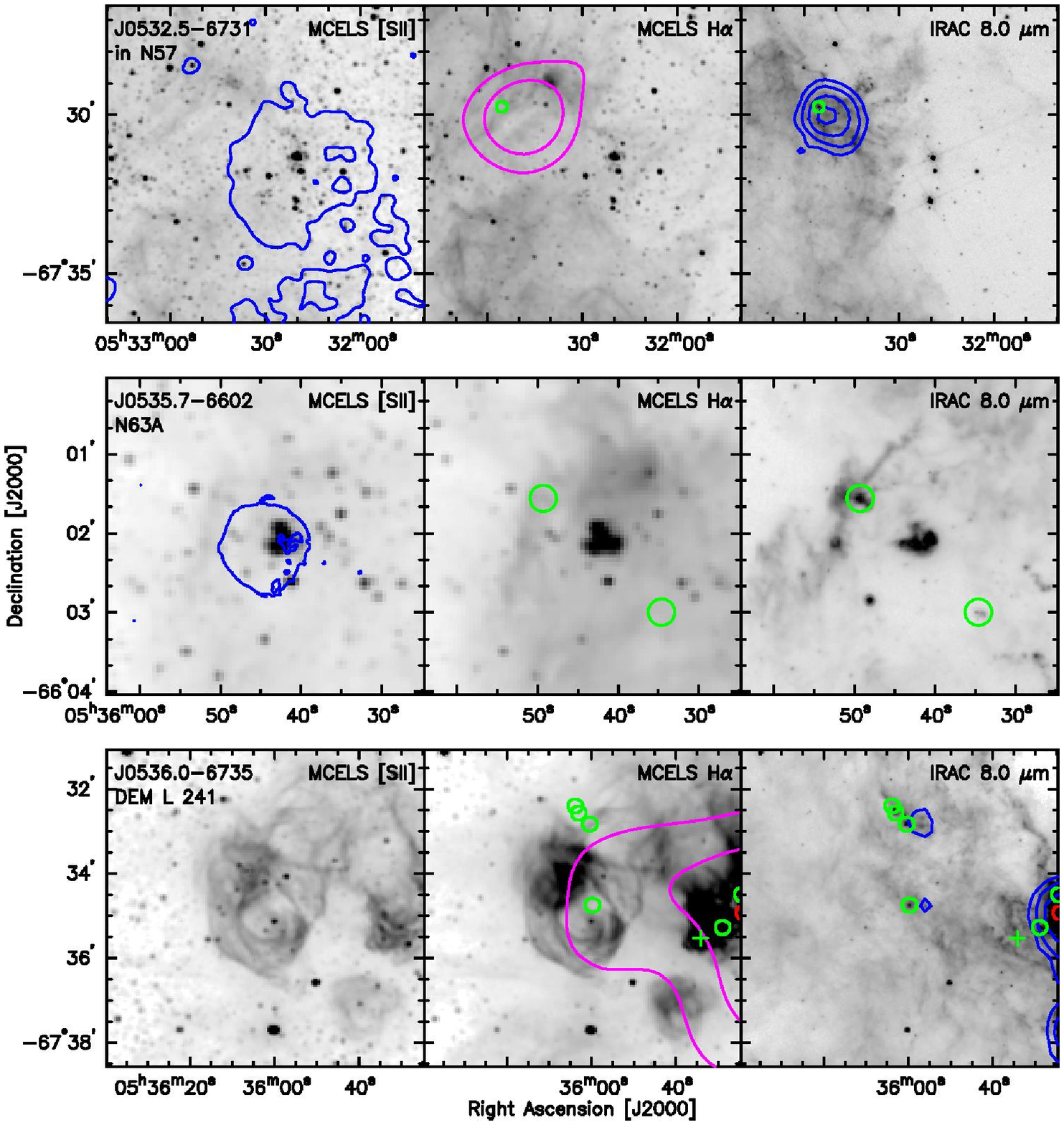}
\end{figure}
\begin{figure}
\plotone{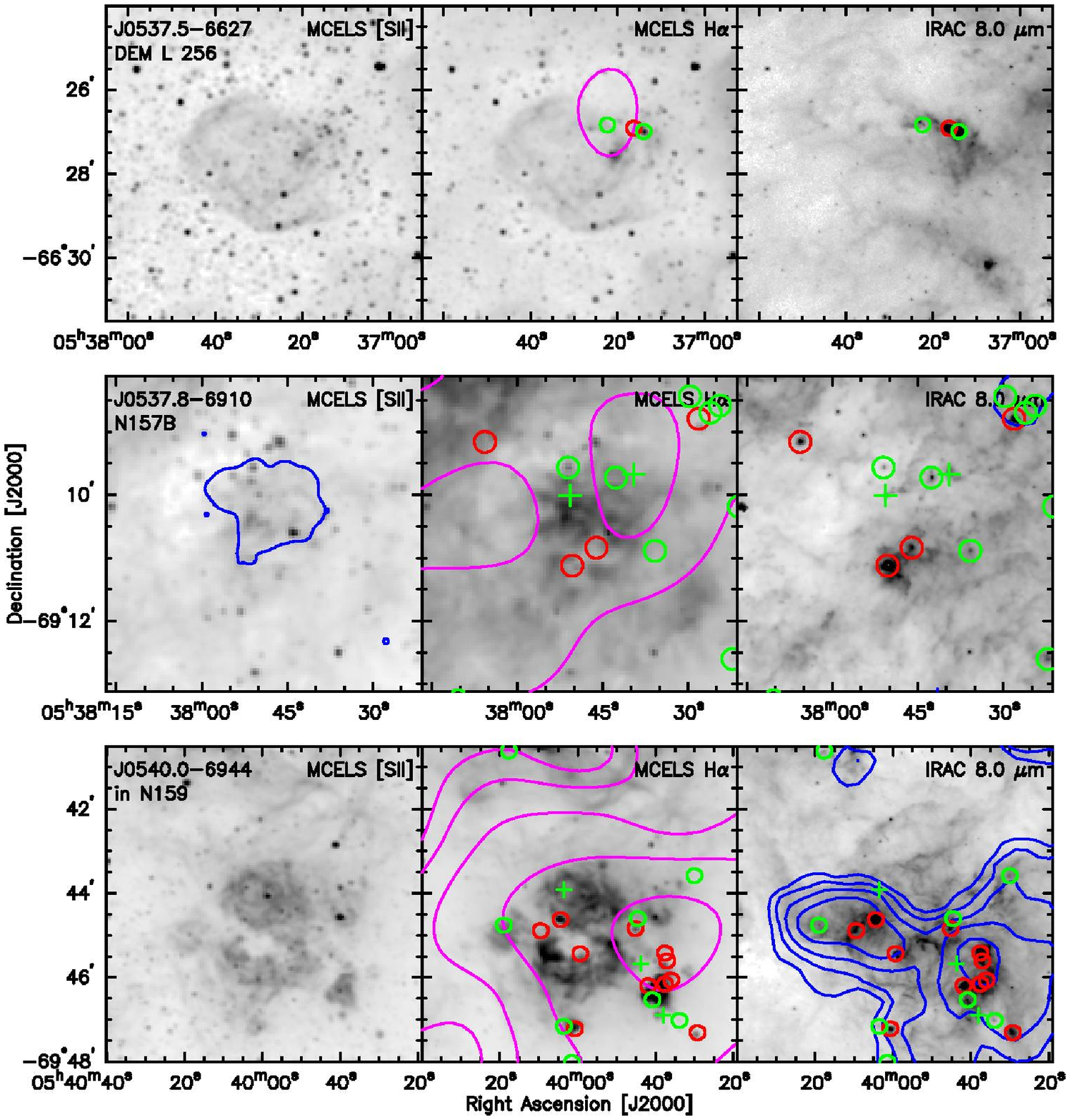}
\end{figure}
\begin{figure}
\plotone{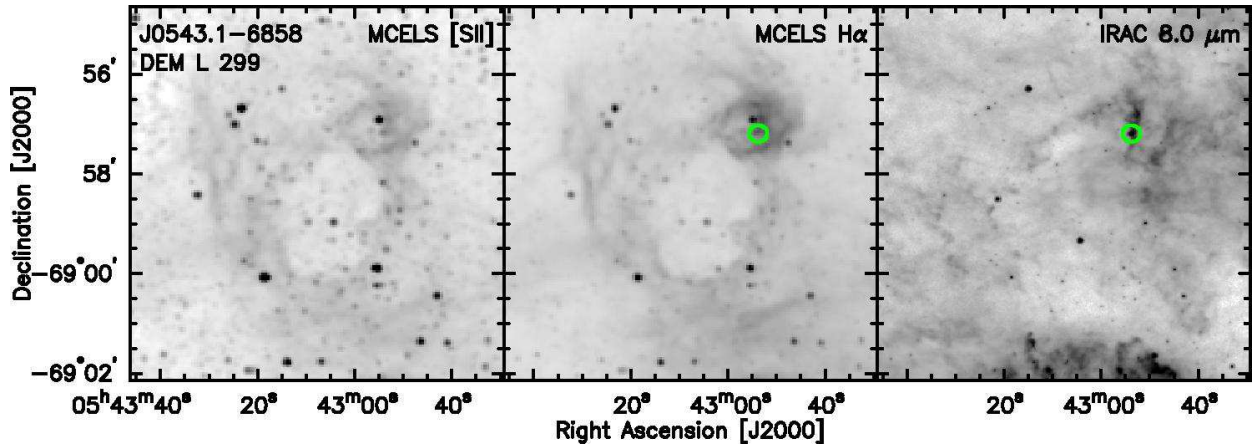}
\caption{
Images of SNRs that are associated with YSOs.  For each remnant the three
panels show images of MCELS [\ion{S}{2}] (left), MCELS H$\alpha$ (middle),
and {\it Spitzer} IRAC 8.0~$\mu$m (right) emission.  The emission line images
are not continuum subtracted.  Definite and probable YSOs from GC09 are
marked with open circles while possible YSOs are indicated by a cross.
Red symbols are used to mark the positions of massive YSOs while green
symbols mark the intermediate-mass YSOs. Contours showing
CO(J=1$\rightarrow$0) emission from the NANTEN survey are overlaid 
on the H$\alpha$
images while contours from the MAGMA survey are overlaid on the
IRAC 8.0 $\mu$m images.  In both cases the contour levels used reflect
the typical RMS noise level, $\sigma$, for the integrated intensity maps
(0.3~K~km~s$^{-1}$ and 0.5~K~km~s$^{-1}$ for NANTEN and MAGMA,
respectively) and follow a geometric progression starting at the 3$\sigma$
level (i.e., 3$\sigma$, 6$\sigma$, 12$\sigma$, 24$\sigma$, ...).
Note that MAGMA observations are available only for regions where 
GMCs are detected in the NANTEN survey.
In three cases, J0532$-$6731, J0535.7$-$6602, and J0537.8$-$6910,
we have overlaid a single X-ray contour on the [\ion{S}{2}] image to 
show the extent of the remnant.
}
\end{figure}

\eject

\begin{figure}
\plotone{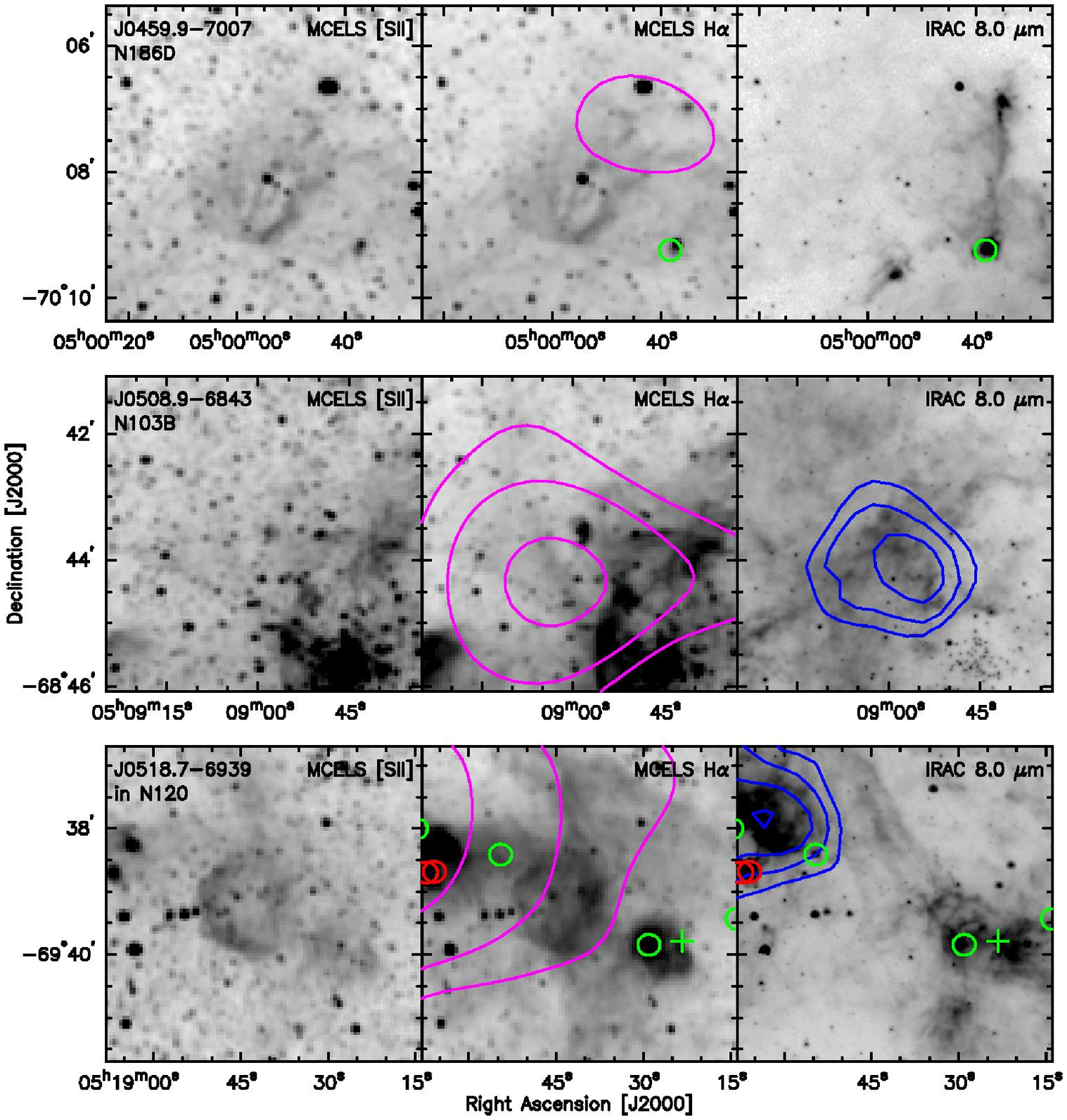}
\end{figure}
\begin{figure}
\plotone{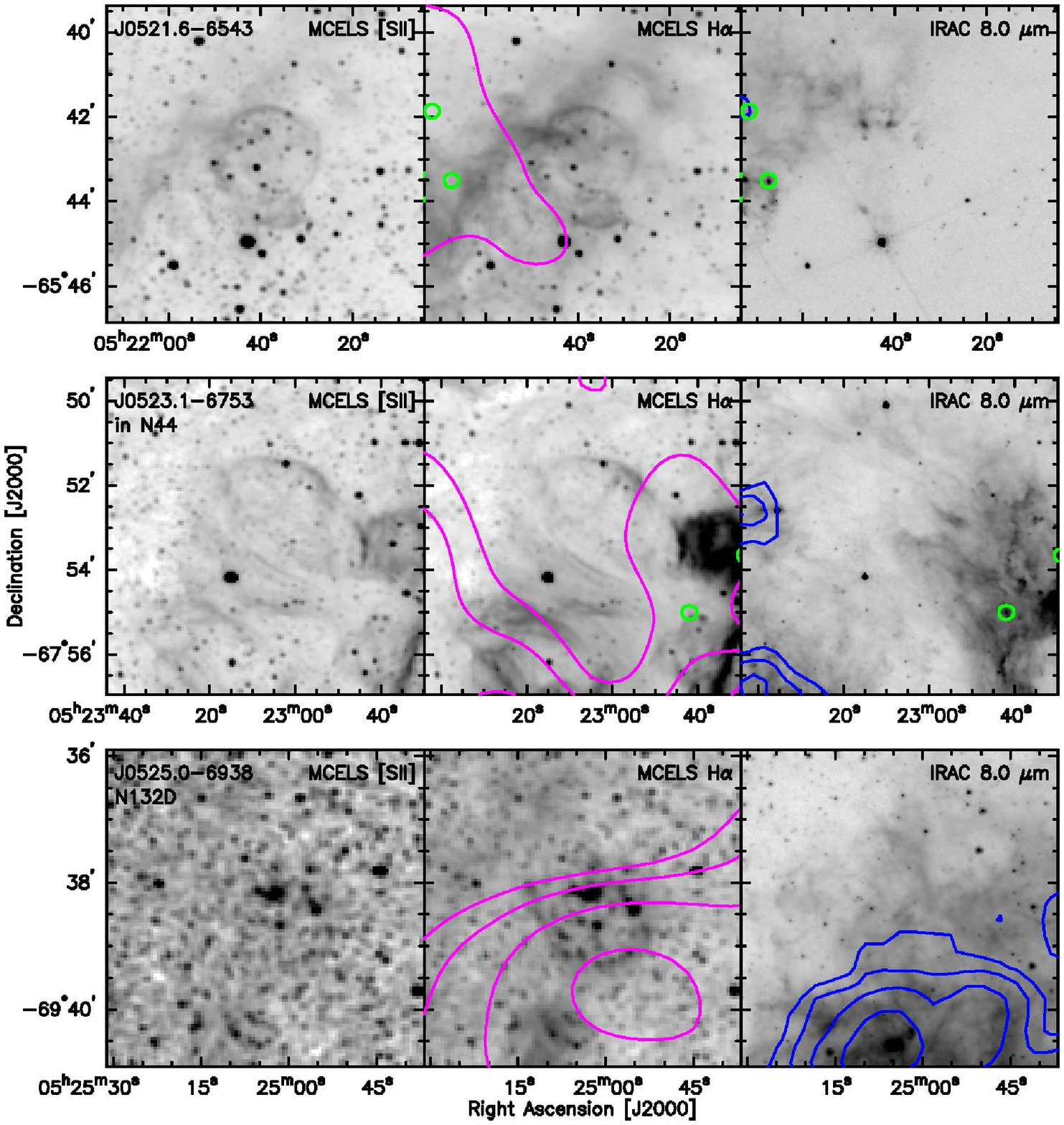}
\end{figure}
\begin{figure}
\plotone{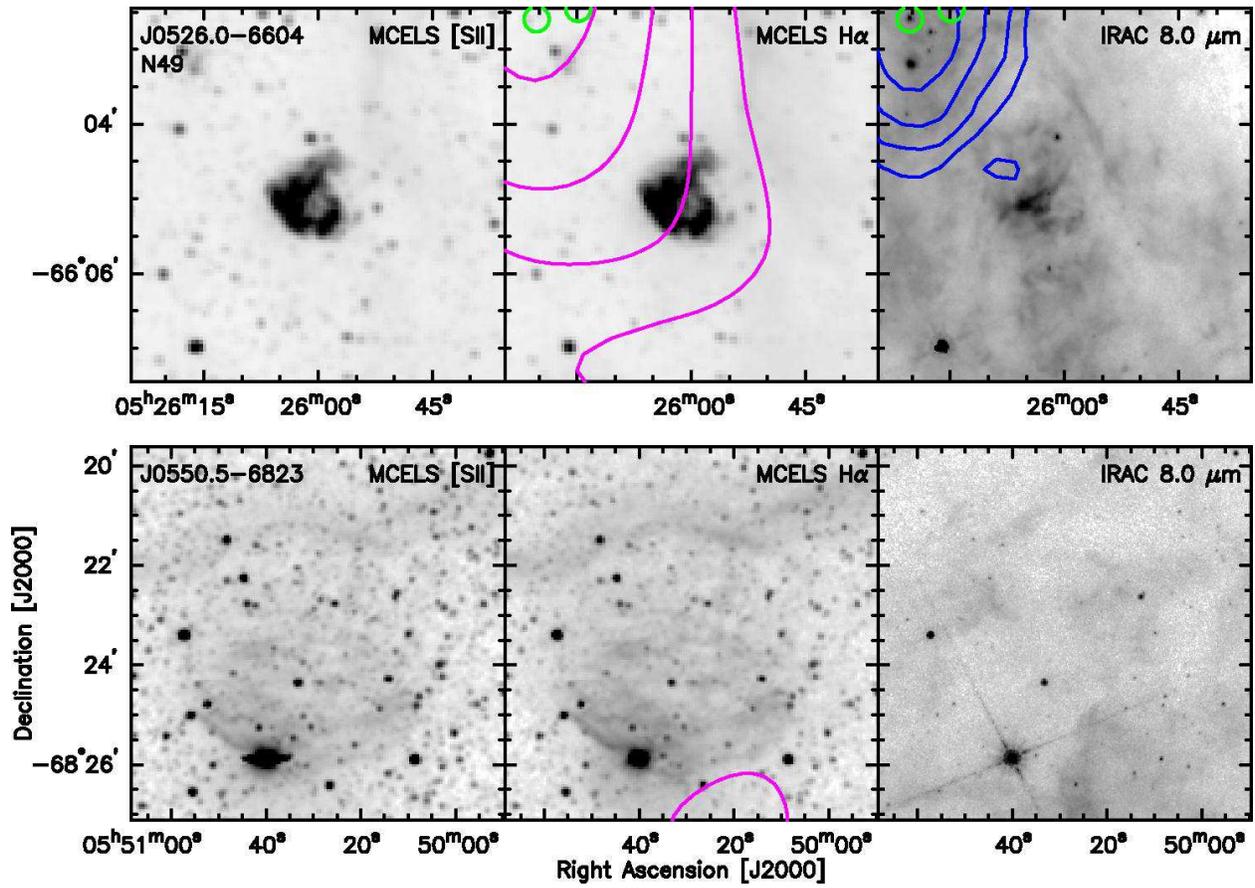}
\caption{
Images of SNRs that are associated with molecular clouds but no YSOs.
See Figure 1 for a description of the contours.
}
\end{figure}

\begin{figure}
\plotone{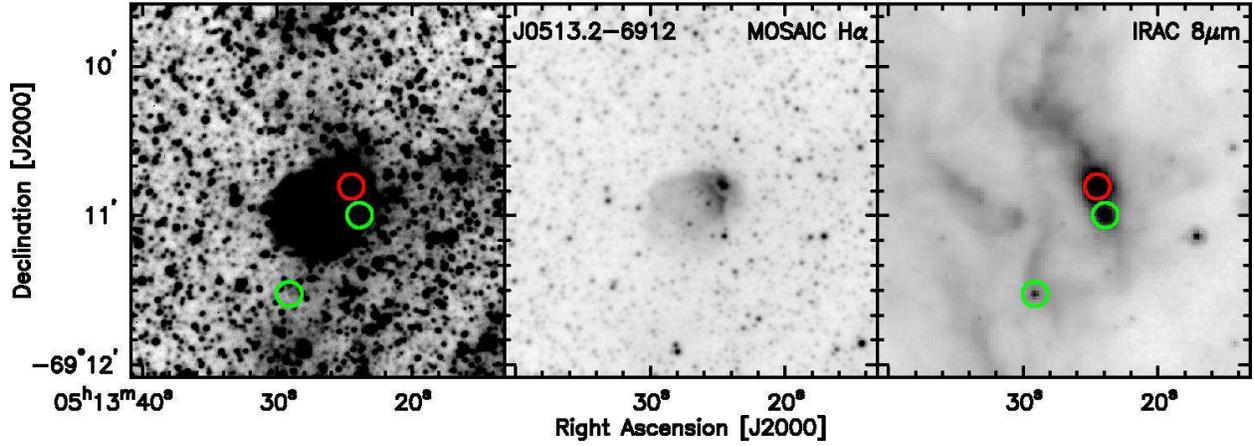}
\caption{Close-up of the northeast rim of SNR J0513.2$-$6912 where 
the current YSOs and a small \HII\ region are located.  The left 
and central panels show an H$\alpha$ image obtained with the MOSAIC 
camera on the CTIO 4m telescope.  The left panel has a linear 
greyscale set to show the relation between the \HII\ region and 
SNR shell rim, while central panel shows the details of the \HII\
region structure.  The right panel shows the 8~$\mu$m emission.  
Locations of massive and intermediate-mass YSOs are marked with 
red and green open circles, respectively.}
\end{figure}

\begin{figure}
\plotone{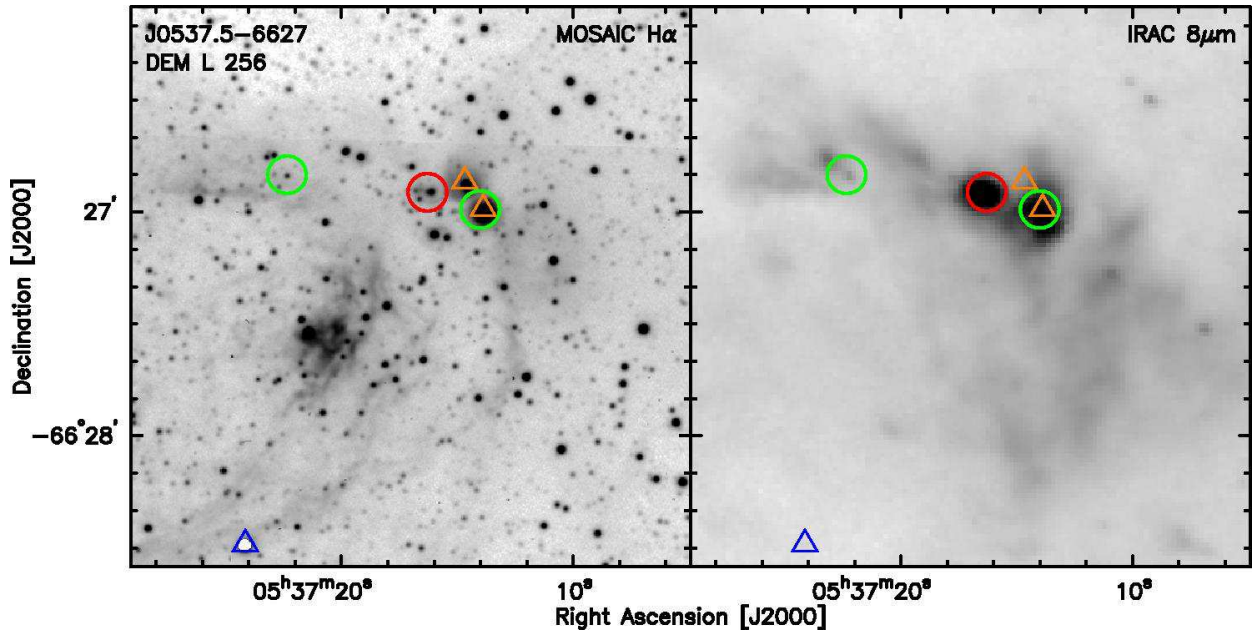}
\caption{
Close-up of the northwest rim of SNR J0537.5$-$6627, where the 
current YSOs are located.  The left panel shows an H$\alpha$ 
image acquired with the MOSAIC camera on the CTIO 4m telescope, 
while the right panel shows the same region at 8~\um.  The red 
and green open circles mark the locations of YSOs already shown 
in Figure~1.  The locations of an O-type star and two early 
B-type stars are marked by blue and orange triangles, respectively.}
\end{figure}

\begin{figure}
\plotone{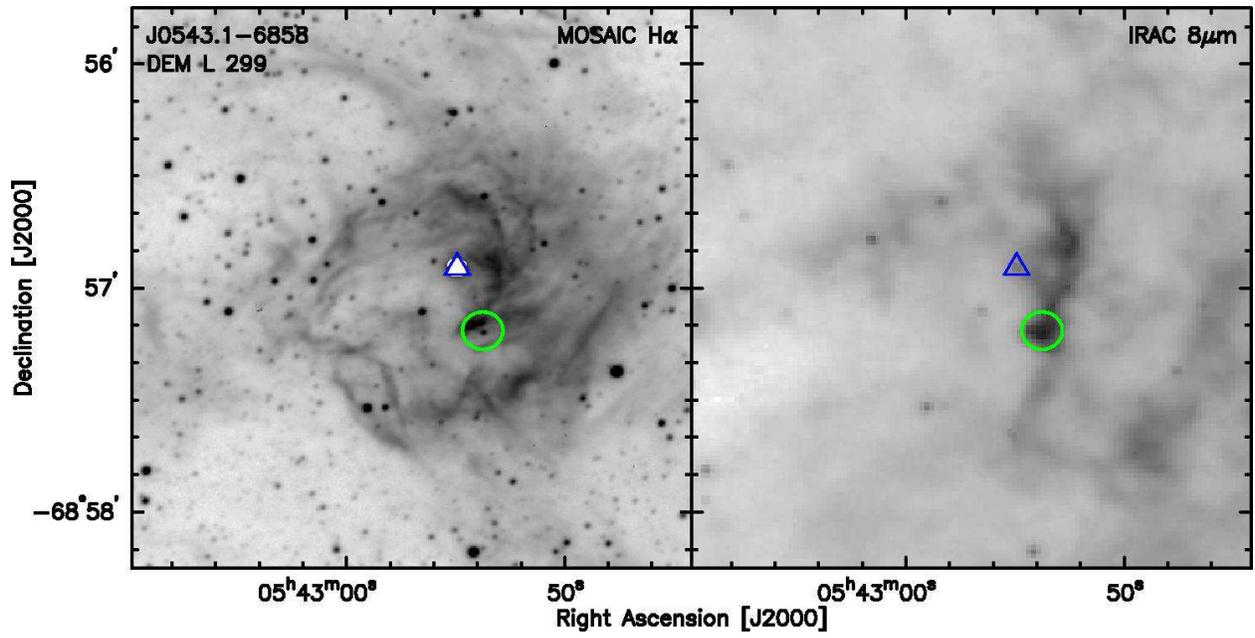}
\caption{
Close-up of the \HII\ region on the northwest rim of SNR J0543.1$-$6858.
The left panel shows an H$\alpha$ image acquired with the MOSAIC camera 
on the CTIO 4m telescope, while the right panel shows the same region 
at 8~\um.  The open circle marks the location of the YSO, while the 
triangle marks the position of Sk~$-$68~155 (saturated in the
H$\alpha$ image).}
\end{figure}

\begin{deluxetable}{lllllccc}
%\tablecolumns{4}
\tabletypesize{\scriptsize}
\tablewidth{0pc}
\tablecaption{Supernova Remnants in the Large Magellanic Cloud}
\tablehead{
SNR         & SNR       & Other       & RA (J2000) & Dec (J2000) &  Size            &  YSOs &  NANTEN  \\
J2000       & B1950     & Name        & (hh mm ss) & (ddd mm ss) &  (arcmin)        &       &  CO      }
\startdata
0448.4$-$6700 & 0448$-$67.1 & MCELS J0448$-$6659& 04 48 25 & $-$67 00 12 & 4.5 $\times$ 3.4   &  N     & N   \\
0449.3$-$6920 & 0449$-$69.4 & MCELS J0449$-$6921& 04 49 22 & $-$69 20 25 &   2.0              &  Y     & N   \\
0450.5$-$7050 & 0450$-$70.9 & SNR 0450$-$709    & 04 50 30 & $-$70 50 05 & 7.7 $\times$ 5.3   &  N     & N   \\
0453.2$-$6655 & 0453$-$66.9 & SNR in N4 	& 04 53 14 & $-$66 55 42 &   4.3              &  N     & N   \\
0453.6$-$6829 & 0453$-$68.5 & SNR 0453$-$685 	& 04 53 38 & $-$68 29 27 &   2.0              &  N     & N   \\	
0454.5$-$6713 & 0454$-$67.2 & SNR in N9 	& 04 54 33 & $-$67 13 00 & 2.8 $\times$ 2.2   &  N     & N   \\
0454.8$-$6625 & 0454$-$66.5 & N11L 	        & 04 54 50 & $-$66 25 37 & 1.4 $\times$ 1.0   &  N     & N   \\
0455.6$-$6838 & 0455$-$68.7 & N86 	        & 04 55 44 & $-$68 38 23 & 6.5 $\times$ 3.5   &  Y     & Y   \\
0459.9$-$7007 & 0500$-$70.2 & N186D 	        & 04 59 56 & $-$70 07 58 & 2.6 $\times$ 2.3   &  N     & Y   \\
0505.7$-$6752 & 0505$-$67.9 & DEM\,L71 	        & 05 05 42 & $-$67 52 39 & 1.5 $\times$ 1.2   &  N     & N   \\
0505.9$-$6801 & 0506$-$68.0 & N23 	        & 05 05 55 & $-$68 01 47 & 1.2 $\times$ 0.8   &  N     & N   \\	
0506.1$-$6541 & 0506$-$65.8 & DEM\,L72          & 05 06 06 & $-$65 41 08 & 6.4 $\times$ 4.7   &  N     & N   \\
0508.9$-$6843 & 0509$-$68.7 & N103B 	        & 05 08 59 & $-$68 43 35 &   0.50             &  N     & Y   \\	
0509.5$-$6731 & 0509$-$67.5 & SNR 0509$-$675 	& 05 09 31 & $-$67 31 17 &   0.56             &  N     & N   \\
0513.2$-$6912 & 0513$-$69.2 & SNR 0513$-$692 	& 05 13 14 & $-$69 12 08 & 4.5  $\times$ 3.2  &  Y     & Y   \\
0518.7$-$6939 & 0519$-$69.7 & SNR in N120 	& 05 18 44 & $-$69 39 09 & 1.6  $\times$ 1.3  &  N     & Y   \\
0519.5$-$6902 & 0519$-$69.0 & SNR 0519$-$690 	& 05 19 35 & $-$69 02 09 &   0.55             &  N     & N   \\
0519.7$-$6926 & 0520$-$69.4 & SNR 0520$-$694 	& 05 19 46 & $-$69 26 00 & 2.4 $\times$ 2.1   &  N     & N   \\
0521.6$-$6543 & 0522$-$65.8 & MCELS J0521$-$6542& 05 21 39 & $-$65 43 10 & 3.0 $\times$ 2.4   &  N     & Y   \\
0523.1$-$6753 & 0523$-$67.9 & SNR in N44 	& 05 23 06 & $-$67 53 06 &   3.5              &  N     & Y   \\
0524.3$-$6624 & 0524$-$66.4 & DEM\,L175a 	& 05 24 18 & $-$66 24 19 & 4.1 $\times$ 2.8   &  N     & N   \\
0525.0$-$6938 & 0525$-$69.6 & N132D 	        & 05 25 04 & $-$69 38 28 & 2.0 $\times$ 1.5   &  N     & Y   \\
0525.4$-$6559 & 0525$-$66.0 & N49B 	        & 05 25 25 & $-$65 59 22 & 2.5 $\times$ 2.3   &  N     & N   \\
0526.0$-$6604 & 0525$-$66.1 & N49 	        & 05 26 00 & $-$66 04 57 & 1.5 $\times$ 1.3   &  N     & Y   \\
0527.6$-$6912 & 0528$-$69.2 & SNR 0528$-$692 	& 05 27 39 & $-$69 12 04 & 2.7 $\times$ 2.0   &  N     & N   \\
0527.9$-$6550 & 0527$-$65.8 & DEM\,L204 	& 05 27 57 & $-$65 50 00 &   4.5              &  N     & N   \\
0530.7$-$7007 & 0531$-$70.2 & MCELS J0530$-$7008& 05 30 44 & $-$70 07 10 & 3.5 $\times$ 2.8   &  N     & N   \\
0531.9$-$7100 & 0532$-$71.0 & SNR in N206 	& 05 31 56 & $-$71 00 19 &   3.0              &  N     & N   \\
0532.5$-$6731 & 0532$-$67.5 & SNR 0532$-$675 in N57 & 05 32 30 & $-$67 31 33 & 4.5            &  Y     & Y   \\
0534.0$-$6955 & 0534$-$69.9 & SNR 0534$-$699 	& 05 34 02 & $-$69 55 05 & 1.7 $\times$ 1.4   &  N     & N   \\
0534.3$-$7033 & 0534$-$70.5 & DEM\,L238 	& 05 34 18 & $-$70 33 26 & 2.9 $\times$ 2.5   &  N     & N   \\
0535.4$-$6916 & 0535$-$69.3 & SNR 1987A 	& 05 35 28 & $-$69 16 11 & $<$ 0.1            &  N     & N   \\
0535.7$-$6602 & 0535$-$66.0 & N63A 	        & 05 35 44 & $-$66 02 14 & 1.4 $\times$ 1.2   &  Y     & N   \\
0535.8$-$6918 & 0536$-$69.3 & Honeycomb 	& 05 35 48 & $-$69 18 04 & 1.4 $\times$ 0.6   &  N     & N   \\
0536.0$-$6735 & 0536$-$67.6 & DEM\,L241 	& 05 36 03 & $-$67 35 04 &   2.4              &  Y     & Y   \\
0536.1$-$7038 & 0536$-$70.6 & DEM\,L249 	& 05 36 07 & $-$70 38 37 & 3.0 $\times$ 2.0   &  N     & N   \\
0537.5$-$6627 & 0538$-$66.5 & DEM\,L256         & 05 37 30 & $-$66 27 47 & 3.6 $\times$ 2.8   &  Y     & Y   \\
0537.8$-$6910 & 0538$-$69.1 & N157B 	        & 05 37 48 & $-$69 10 35 & 1.7 $\times$ 1.2   &  Y     & Y   \\
0540.0$-$6944 & 0540$-$69.7 & SNR in N159 	& 05 40 00 & $-$69 44 02 &   1.8              &  Y     & Y   \\
0540.2$-$6919 & 0540$-$69.3 & N158A       	& 05 40 12 & $-$69 19 55 & 1.3 $\times$ 1.1   &  N     & N   \\
0543.1$-$6858 & 0543$-$68.9 & DEM\,L299 	& 05 43 10 & $-$68 58 49 & 5.8 $\times$ 4.0   &  Y     & N   \\
0547.0$-$6942 & 0547$-$69.7 & DEM\,L316 B 	& 05 47 00 & $-$69 42 55 & 3.4 $\times$ 2.8   &  N     & N   \\
0547.3$-$6941 & 0547$-$69.7 & DEM\,L316 A 	& 05 47 22 & $-$69 41 26 &   2.0              &  N     & N   \\
0547.8$-$7024 & 0548$-$70.4 & SNR 0548$-$704 	& 05 47 49 & $-$70 24 52 & 2.0 $\times$ 1.8   &  N     & N   \\
0550.5$-$6823 & 0551$-$68.4 & SNR J0550$-$6823  & 05 50 30 & $-$68 23 22 & 5.2 $\times$ 3.5   &  N     & Y   
\enddata
\end{deluxetable}

\end{document}